\documentclass[fleqn,12pt,twoside]{article}
\usepackage[headings]{espcrc1}
\readRCS
$Id: espcrc1.tex,v 1.2 2004/02/24 11:22:11 spepping Exp $
\usepackage{graphicx}

\newcommand{\AmS}{{\protect\the\textfont2
  A\kern-.1667em\lower.5ex\hbox{M}\kern-.125emS}}

\title{Transverse Momentum Distribution as a Probe of $J/\psi$ Production Mechanism in Heavy Ion Collisions}

\author{Kai Zhou\address[THU]{Physics Department, Tsinghua University, Beijing
100084, China},
        Nu Xu\address{Nuclear Science Division, Lawrence Berkeley
National Laboratory, Berkeley, California 94720, USA}
        and
        Pengfei Zhuang\addressmark[THU]}
\runtitle{}
\runauthor{}
\begin{document}
%
\maketitle
\begin{abstract}
We investigate $J/\psi$ transverse momentum distribution in a
transport approach. While the nuclear modification factor
$R_{AA}(N_p)$ at RHIC is almost the same as at SPS, the averaged
transverse momentum square $\langle p_t^2\rangle$ and $R_{AA}(p_t)$
are very different at SPS, RHIC and LHC and can be used to
differentiate from the $J/\psi$ production mechanisms in high energy
nuclear collisions.
\end{abstract}

\section{Introduction}
From lattice Quantum Chromodynamics (QCD) calculations, there exists
a phase transition from ordinary hadronic matter to a new state of
matter, the so-called Quark Gluon Plasma (QGP), at finite
temperature. The $J/\Psi$ suppression has long been considered as a
probe of the new state produced in high energy heavy ion
collisions~\cite{matsui}. The primordially produced charmonia via
hard nucleon-nucleon (NN) collisions are subject to subsequent
nuclear absorption in the initial stage and anomalous suppression in
the hot and dense medium. The normal and anomalous suppression are
indeed observed in heavy ion collisions at the CERN Super Proton
Synchrotron (SPS)~\cite{sps1} and investigated in many theoretical
models~\cite{spsreview}.

Different from the $J/\psi$ production at SPS, there are a
remarkable number of charm quarks in the QGP phase produced in
higher energy nuclear collisions at the BNL Relativistic Heavy Ion
Collider (RHIC) and the CERN Large Hadron Collider (LHC), and the
regeneration, namely the recombination of those uncorrelated charm
quarks offers another source for $J/\psi$
production~\cite{regeneration}. Obviously, the regeneration will
enhance the $J/\psi$ yield and alter its momentum spectrum.

From recently observed $J/\psi$ production at RHIC~\cite{rhic1}, the
nuclear modification factor $R_{AA}(N_p)$ as a function of the
number of participant nucleons $N_p$ at RHIC is almost the same as
at SPS, see Fig.\ref{fig1}. The same suppression at SPS and RHIC
looks difficult to be understood in models with only initial
production mechanism, because the temperature at RHIC is higher and
then the anomalous suppression is predicted to be stronger at RHIC,
in comparison with SPS. The puzzle of the same suppression was
theoretically studied by many models. With the idea of sequential
suppression~\cite{satz}, if the temperatures at RHIC and SPS are
both in between the $J/\psi$ dissociation temperature and $\psi'$
and $\chi_c$ dissociation temperature, the value $R_{AA}\sim 0.6$
will not change from SPS to RHIC energy. Considering three gluon
fusion as the main $J/\psi$ production mechanism~\cite{kharzeev},
the cold nuclear matter effect which is medium independent can
explain the same suppression too. In the frame of regeneration, the
competition between the initial production and regeneration can
explain well the same suppression~\cite{zhao,liu}.

The transverse momentum distribution contains more dynamic
information on the charmonium production and suppression mechanism.
The regenerated $J/\psi$s are mainly distributed in low $p_t$ and
central rapidity region~\cite{liu}, but the high $p_t$ region is
closely related to the Cronin effect~\cite{gavin} and leakage
effect~\cite{hufner} for the initially produced $J/\psi$s. While the
$J/\psi$ yield which is a global quantity is not sensitive to the
detailed dynamics, the transverse momentum distribution at RHIC is
very different from SPS, see Fig.\ref{fig2}, and may be used to
differentiate from the production and suppression mechanisms at
different energies. In this paper, we investigate the $J/\psi$
transverse momentum moments and the $p_t$ dependence of $R_{AA}$
from SPS to LHC energy.

\section{Transport Model and Numerical Results}
We start with a transport equation~\cite{zhu,liu} for the
distribution function $f_\Psi({\bf p}_t,{\bf x}_t,\tau|{\bf b})$ in
central rapidity region and transverse phase space $({\bf p}_t,{\bf
x}_t)$ at time $\tau$ and fixed impact parameter ${\bf b}$,
\begin{equation}
\label{transport}
\partial f_\Psi/\partial \tau +{\bf
v}_\Psi\cdot{\bf \nabla}f_\Psi = -\alpha_\Psi f_\Psi +\beta_\Psi,
\end{equation}
where $\Psi$ stands for $J/\psi$, $\psi'$ and $\chi_c$, ${\bf
v}_\Psi = {\bf p}_t/\sqrt{p_t^2+m_\Psi ^2}$ is the transverse
velocity, and $\alpha_\Psi({\bf p}_t,{\bf x}_t,\tau|{\bf b})$ and
$\beta_\Psi({\bf p}_t,{\bf x}_t,\tau|{\bf b})$ are the loss and gain
terms representing the anomalous suppression and regeneration in the
hot medium. Considering the gluon dissociation process $g+\Psi\to
c+\bar c$, $\alpha$ is the momentum integration of the dissociation
cross section~\cite{peskin} multiplied by thermal gluon
distribution, and $\beta$ can be obtained from detailed balance
between the suppression and regeneration. The distribution
$f_\Psi({\bf p}_t,{\bf x}_t,\tau_0|{\bf b})$ at initial time
$\tau_0$ is determined by the geometrical superposition of NN
collisions, including the Cronin effect~\cite{gavin} and nuclear
absorption~\cite{spsreview}. The local temperature, baryon chemical
potential and collective flow appeared in the thermal gluon and
charm quark distribution functions are determined by the ideal
hydrodynamic equations~\cite{zhu}.

In comparison with the QGP phase, the particle density in the
hadronic phase which appears in the later period of the system
evolution is much lower. To simplify the numerical calculation, we
neglect the hadron contribution to the $J/\psi$ production. Solving
the coupled transport equation for the charmonium motion and the
hydrodynamic equations for the QGP evolution, one can obtain the
$J/\psi$ distribution function at the hadronization time and then
get the final state $J/\psi$ yield and transverse momentum
distribution.

\begin{figure}[!htb]\centering
\vspace{-1cm}
\includegraphics[height=7cm]{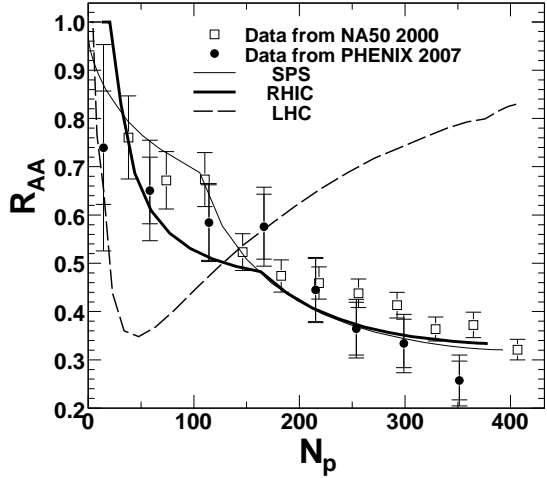}%
\vspace{-1cm} \caption{The nuclear modification factor $R_{AA}$ as a
function of $N_p$ for Pb+Pb collisions at SPS and LHC and Au+Au
collisions at RHIC at mid rapidity. The SPS and RHIC data are from
\cite{sps1} and \cite{rhic1}, and the lines are our theoretical
calculations. } \label{fig1} \vspace{-0.5cm}
\end{figure}

We now calculate the nuclear modification factor $R_{AA}$ and
averaged transverse momentum square $\langle p_t^2\rangle$ as
functions of the number of participant nucleons $N_p$ for $J/\psi$s
produced in Pb+Pb collisions at SPS and LHC energy and Au+Au
collisions at RHIC. All the calculations are in mid rapidity. The
corresponding parameters for the initial charmonium and charm quark
distributions in NN collisions and for the hot medium can be found
in Ref.\cite{liu}. At RHIC, both the suppression and regeneration in
the medium are stronger than at SPS, the competition between the
suppression and regeneration leads to almost the same $R_{AA}$ at
RHIC and SPS, especially for semi-central and central collisions
with $N_p>150$, see Fig.\ref{fig1}. However, the case is very
different at LHC. The initially produced $J/\psi$s are almost all
eaten up by the very hot, long lived and large fireball, and the
regeneration becomes dominant in a wide region of $N_p$. Only for
peripheral collisions with $N_p < 50$, the initial production is
important. Due to the increasing suppression and regeneration with
centrality, the $R_{AA}$ decreases with $N_p$ in the initial
production dominant region and increases with $N_p$ in the
regeneration dominant region, see Fig.\ref{fig1}.

\begin{figure}[!htb]
\vspace{-1cm}
\includegraphics[height=6.35cm]{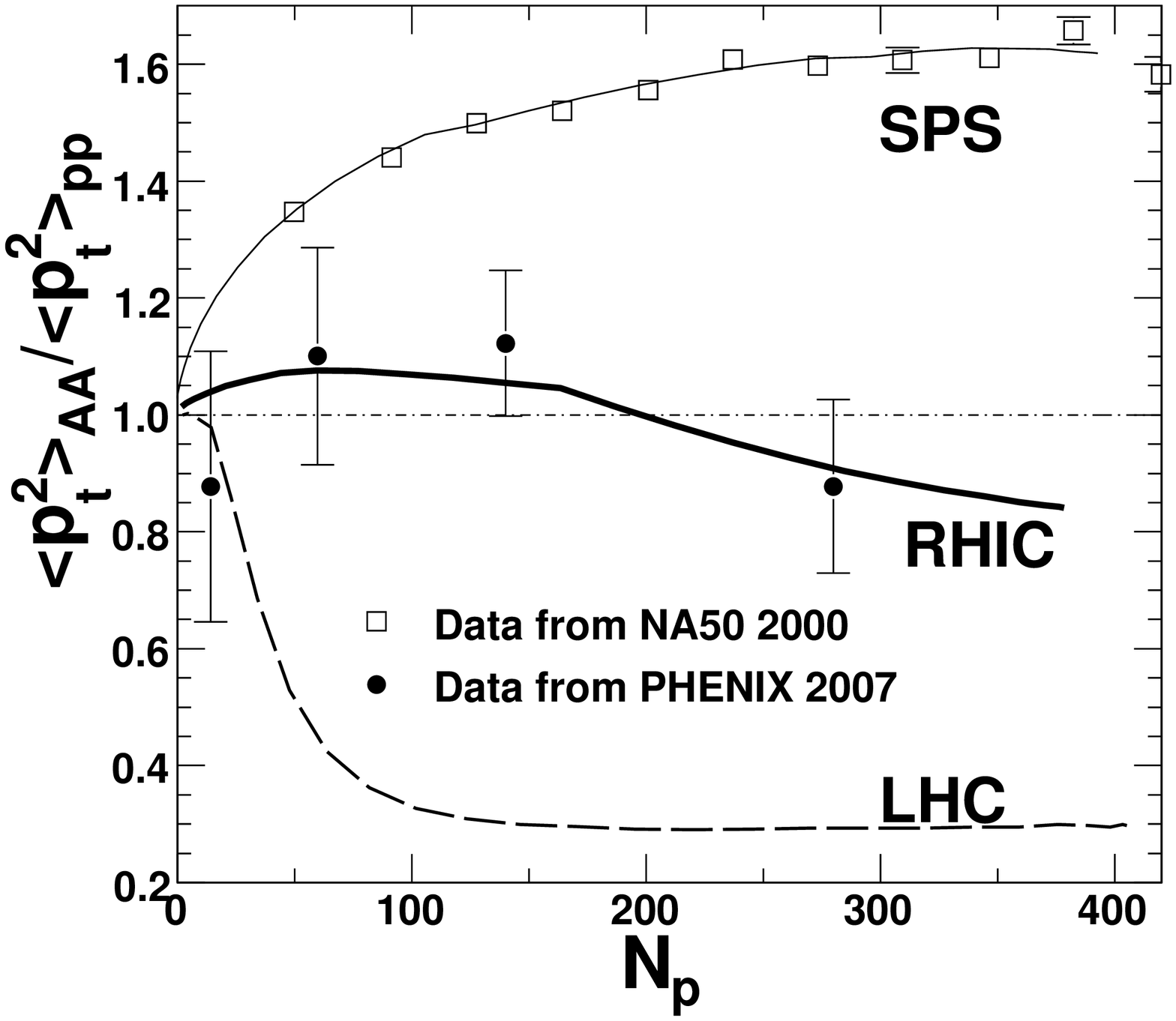}%
\hspace{1cm}%
\includegraphics[height=6.5cm]{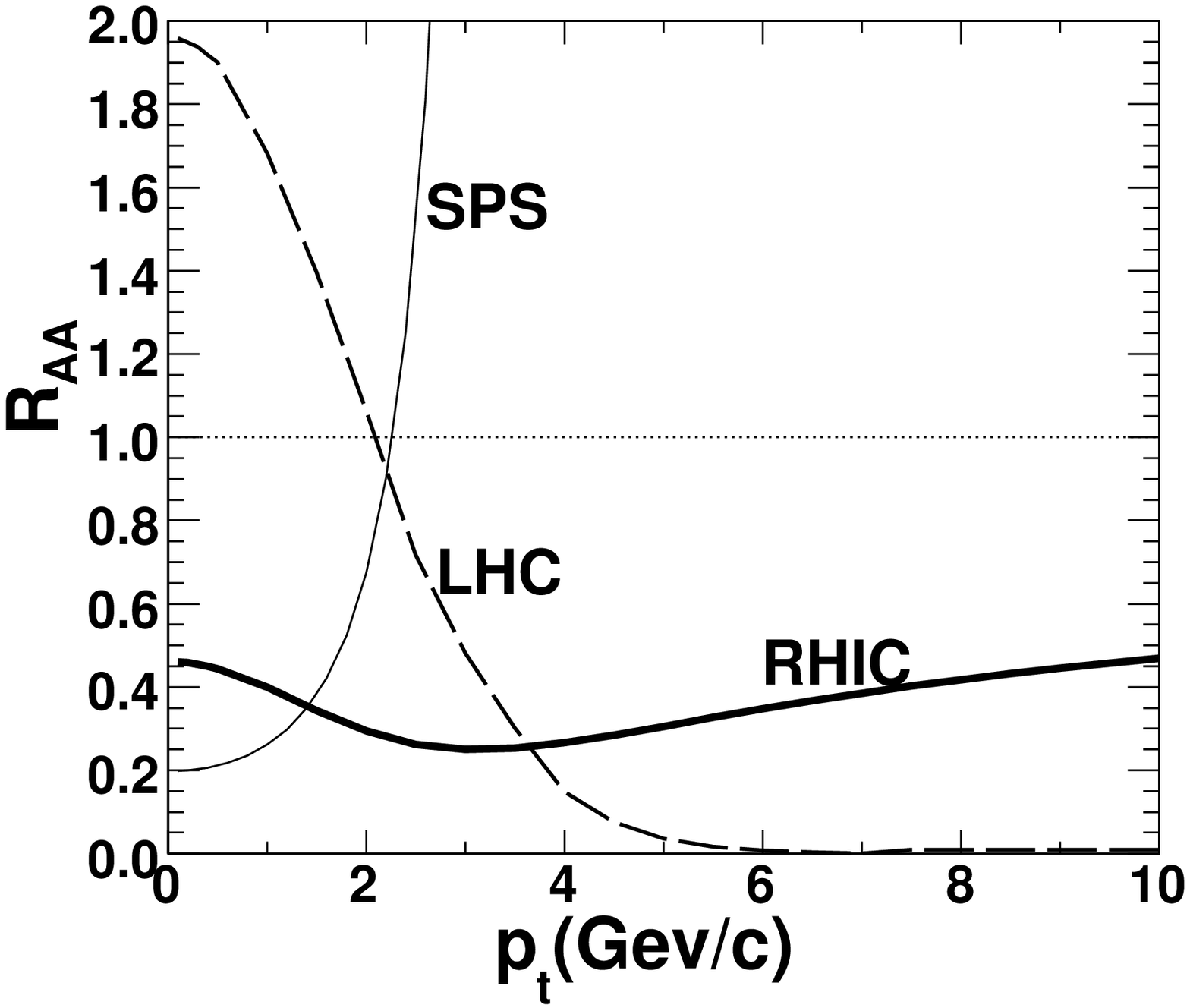}
\vspace{-1cm} \caption{The normalized averaged transverse momentum
square $\langle p_t^2\rangle_{AA}/\langle p_t^2\rangle_{pp}$ as a
function of $N_p$ (left panel) and $R_{AA}$ at $b=0$ as a function
of $p_t$ (right panel) for Pb+Pb collisions at SPS and LHC and Au+Au
collisions at RHIC at mid rapidity. The SPS and RHIC data are from
\cite{sps2} and \cite{rhic2}, and the lines are our theoretical
calculations. } \label{fig2} \vspace{-0.5cm}
\end{figure}

Now we come to the $p_t$ distribution. Fig.\ref{fig2} shows the
averaged transverse momentum square normalized by the corresponding
value in NN collisions, $\langle p_t^2\rangle_{AA}/\langle
p_t^2\rangle_{pp}$, as a function of centrality (left panel) and the
$R_{AA}$ as a function of $p_t$ (right panel). While the
$R_{AA}(N_p)$ is almost the same at SPS and RHIC, the transverse
momentum distribution is really sensitive to the production
mechanism. The dominant initial production, Cronin effect and
leakage effect at SPS lead to an increasing $\langle
p_t^2\rangle_{AA}/\langle p_t^2\rangle_{pp}$ with $N_p$ and an
increasing $R_{AA}(p_t)$ with $p_t$. In contrast to SPS, the
regeneration is the dominant production mechanism at LHC which
results in decreasing $\langle p_t^2\rangle_{AA}/\langle
p_t^2\rangle_{pp}$ and $R_{AA}(p_t)$. Since we assumed charm quark
thermalization in the medium, $\langle p_t^2\rangle_{AA}/\langle
p_t^2\rangle_{pp}$ becomes saturated when the contribution from the
initial production disappears. Considering the fact that the
regenerated $J/\psi$s carry low momentum, the $R_{AA}(p_t)$ is
larger than 1 at low $p_t$ but vanishes at high $p_t$. The case at
RHIC is in between SPS and LHC where both the initial production and
regeneration are important and the competition between them controls
the $J/\psi$ production. It is the contribution from the
regeneration that separates clearly the $J/\psi$ transverse momentum
distribution at RHIC from that at SPS.

In summary, we studied the contribution from the primordial
production in the initial stage and the regeneration in the medium
to the $J/\psi$ production. We found that with increasing fraction
of the regeneration from SPS to LHC, the transverse momentum
distribution behaves very differently and can be used to
differentiate from the $J/\psi$ production mechanism in high energy
nuclear collisions.

\section*{Acknowledgement}
This work is supported by the NSFC Grant 10735040, the 973-project
2007CB815000, and the U.S. Department of Energy under Contract No.
DE-AC03-76SF00098.

\end{document}